\documentclass[aps,prc,twocolumn,superscriptaddress,showpacs,amsmath,floatfix,
nofootinbib]{revtex4}
\usepackage{graphicx}
\usepackage{bm}

\begin{document}


\title{Spin distribution of nuclear levels using static path approximation with
random-phase approximation}

\author{K.~Kaneko}
\email{kaneko@ip.kyusan-u.ac.jp}
\affiliation{Department of Physics, Kyushu Sangyo University, Fukuoka 813-8503,
Japan}
\author{A. Schiller}
\affiliation{National Superconducting Cyclotron Laboratory, Michigan State 
University, East Lansing, MI 48824}

\begin{abstract}
We present a thermal and quantum-mechanical treatment of nuclear rotation using
the formalism of static path approximation (SPA) plus random-phase 
approximation (RPA)\@. Naive perturbation theory fails because of the presence 
of zero-frequency modes due to dynamical symmetry breaking. Such modes lead to 
infrared divergences. We show that composite zero-frequency excitations are 
properly treated within the collective coordinate method. The resulting 
perturbation theory is free from infrared divergences. Without the assumption 
of individual random spin vectors, we derive microscopically the spin 
distribution of the level density. The moment of inertia is thereby related to 
the spin-cutoff parameter in the usual way. Explicit calculations are performed
for $^{56}$Fe; various thermal properties are discussed. In particular, we 
demonstrate that the increase of the moment of inertia with increasing 
temperature is correlated with the suppression of pairing correlations. 
\end{abstract}


\maketitle

\section{Introduction}
\label{sec1}

The spin distribution of nuclear levels is important for Hauser-Feshbach-type 
calculations of astrophysical reaction rates \cite{Rauscher}\@. Generally, the 
spin distribution of nuclear levels is assumed to be given by 
\begin{eqnarray}
\rho(E,I)&=&\rho(E)\frac{2I+1}{2\sigma^2}e^{-\frac{I(I+1)}{2\sigma^2}}
\label{eq:1}\\
&=&W(E)\frac{2I+1}{2\sqrt{2\pi}\sigma^3}e^{-\frac{I(I+1)}{2\sigma^2}},
\label{eq:2}
\end{eqnarray}
where $\rho(E,I)$ is the level density for a given spin $I$ and $\sigma$ is the
spin-cutoff parameter \cite{Bethe,Gilbert,Grimes,Lu,Egidy}\@. It should be 
noted that the total level density (summed over all spins) is given by 
$\rho(E)=\sum_{I}\rho(E,I)$, while the total state density given by 
$W(E)=\sum_{I}(2I+1)\rho(E,I)$ contains an additional factor $2I+1$ for each 
level in order to take into account the $m$ degeneracy. In deriving the spin 
distribution, one typically assumes that the individual nucleon spins are 
pointing in random directions, hence the spin distribution becomes Gaussian and
can be described by only one parameter, the spin cutoff parameter. This 
parameter can be related to the moment of inertia ${\cal J}_{\mathrm{rigid}}$ 
by 
\begin{equation}
\sigma^2=\frac{{\cal J}_{\mathrm{rigid}}T}{\hbar^2},
\end{equation}
where $T$ is the thermodynamical temperature \cite{Ericson}\@. This result can 
be derived within an independent-particle model with individual nucleon spins 
but neglecting correlations \cite{Ericson}\@. Microscopic calculations of the 
spin distribution are difficult when correlations are present. Experimental 
data on the spin-cutoff parameter are available only in a few cases 
\cite{Grimes}\@. Therefore, theoretical predictions of the spin-cutoff 
parameter are greatly desired. Alhassid {\it et al.\@} have recently 
investigated the spin distribution in the framework of the shell model Monte 
Carlo (SMMC) \cite{Alhassid1,Alhassid2} and the static path approximation (SPA)
method \cite{Alhassid1}\@. Assuming that the spin distribution can be described
by Eq.\ (\ref{eq:2}), they found a significant suppression of the moment of 
inertia and an odd-even staggering of the spin-cutoff parameter due to pairing 
correlations at low excitation energies. 

The SPA method \cite{Alhassid1,Alhassid3,Bertsch,Rossignoli1} is a useful 
treatment to evaluate approximately the partition function of finite systems 
with separable interactions such as the pairing plus quadrupole-quadrupole 
(P+QQ) interaction, and it provides the exact result at high temperatures. 
However, as temperatures decrease the SPA becomes inaccurate because quantal 
fluctuations around the mean field can no longer be neglected. In leading order
perturbation theory, small amplitude fluctuations give corrections to the 
partition function which can be treated by random-phase approximation (RPA) 
\cite{Puddu1,Lauritzen,Rossignoli,Attias}\@.

However, such a perturbative treatment exhibits an infrared problem in the case
where the nucleus is modeled in terms of independent nucleons moving in a 
deformed mean field: there are zero-frequency modes among normal vibrational 
modes. This can be understood because the SPA method breaks the symmetry of 
rotational invariance for a deformed nucleus. It operates in a body-fixed 
reference frame which undergoes time-dependent transformations. The collective 
(zero-frequency) excitation, i.e., the motion of the rotating frame, is treated
on an equal footing with other excitations. This leads to a Lagrangian which 
displays local symmetry yielding the mechanical equivalent of gauge invariance.
The collective coordinate method developed by Gervais and Sakita 
\cite{Gervais1} treats the inherent symmetries of the problem consistently. It 
separates the so-called spurious motions from the intrinsic vibrations. In the 
laboratory frame, these zero-frequency excitations are not really spurious; 
they represent a rotational motion which has to be treated separately. 

In the present paper, we describe microscopically the nuclear rotation of a 
deformed nucleus including thermal and quantal fluctuations using the SPA+RPA 
method, and derive the associated spin distribution of nuclear levels. First, 
using the collective coordinate method, the rotation is described in terms of 
the zero-frequency modes which emerge from the breaking of rotational 
invariance of the deformed mean field. Secondly, the spin distribution of 
nuclear levels is calculated by a proper treatment of the zero-frequency modes 
in the framework of the SPA+RPA method. In Sect.\ \ref{sec2}, the collective 
coordinate method is applied to the partition function of an interacting 
fermion system. In Sect.\ \ref{sec3} we show how to calculate the partition 
function of the P+QQ model using the SPA+RPA method. In Sect.\ \ref{sec4}, we 
present a correct treatment of the zero-frequency modes and derive the 
spin-dependent level density. In Sect.\ \ref{sec5}, the model is applied to 
$^{56}$Fe and related thermal properties are investigated. Concluding remarks 
are given in Sect.\ \ref{sec6}\@. 

\section{The collective coordinate method}
\label{sec2}

According to the theory developed by Gervais, Jevicki, and Sakita 
\cite{Gervais2}, a problem involving only fermion degrees of freedom can be 
transformed into an equivalent, constraint problem including both fermion and 
collective degrees of freedom. Concerning the problem at hand, a perturbative 
treatment of nuclear rotation within the path integral formalism has been 
proposed using the collective coordinate method \cite{Alessandrini,Bes}\@. Let 
us consider a partition function of an interacting fermion system given by the 
path integral after Wick rotation 
\begin{equation}
Z=\int\prod_{\alpha}{\cal D}[c^\dagger_\alpha]{\cal D}[c_\alpha]e^{-S_E
(\beta)}.
\label{eq:4}
\end{equation}
Here, the action $S_E(\beta)$ is given by 
\begin{equation}
S_E(\beta)=\int_0^\beta\left(\sum_\alpha c_\alpha^\dagger\dot{c}_\alpha+
H(c^\dagger,c)\right)d\tau,
\end{equation}
where $\beta=1/T$ is the inverse temperature and the dot indicates a derivative
with respect to the imaginary time $\tau$\@. The antiperiodic boundary 
condition $c(\beta)=-c(0)$ arises since the operators $c$ are Grassmann numbers
due to the fermionic nature of the field. The action $S_{E}(\beta)$ is 
rotationally invariant with respect to the SO(3) algebra. 

We shall now be concerned with the case in which the fermion system becomes 
deformed and breaks the SO(3) invariance. As mentioned in the Introduction, a 
naive perturbative treatment fails due to the presence of zero-frequency modes.
A better way to tackle this problem is to introduce collective coordinates. We 
therefore adopt the path integral formulation of Gervais and Sakita 
\cite{Gervais1}\@. In order to separate the collective coordinates, we employ 
the equality 
\begin{eqnarray}
1&=&\int\prod_i{\cal D}[L_i]{\cal D}[\phi_i]\delta(\hbar L_i-J_i(b^\dagger,b))
\nonumber\\
&&\times\delta(\Theta_i(b^\dagger,b))\det[J,\Theta]_{\mathrm{PB}},
\label{eq:6}
\end{eqnarray}
where $\Theta_i(b^\dagger,b)$ is an arbitrary function of the fermion fields 
$(b^\dagger,b)$ with the condition that the determinant of the Poisson bracket 
$[\ ]_{\mathrm{PB}}$ does not vanish. The fermion fields in the body-fixed 
frame $(b^\dagger,b)$ are obtained by a linear transformation from the 
laboratory frame 
\begin{equation}
b_\alpha(\phi_i)=P_{\mathrm{R}}(\phi_i)c_\alpha P_{\mathrm{R}}(\phi_i)^{-1},
\end{equation}
where $P_{\mathrm{R}}(\phi_i)$ is the rotational operator with the Euler angles
$\phi_i\,(i=1,2,3)$, and $L_i$ are the three components of the collective 
angular momentum. 

Inserting the Eq.\ (\ref{eq:6}) into the right hand side of Eq.\ (\ref{eq:4}) 
and exploiting the constraint $\hbar L_i=J_i(b^\dagger,b)$, one finds 
\begin{eqnarray}
\lefteqn{Z=\int\prod_\alpha{\cal D}[b^\dagger_\alpha]{\cal D}
[b_\alpha]\int\prod_i{\cal D}[L_i]{\cal D}[\phi_i]}\nonumber\\
&&\hspace*{-5mm}\times\exp\left(-S_E(\beta)+\int_0^\beta\sum_j\Pi_j(\tau)
\dot{\phi}_j(\tau)d\tau\right)\nonumber\\
&&\hspace*{-5mm}\times\delta(\hbar L_i-J_i(b^\dagger,b))\delta(\Theta_i
(b^\dagger,b))\det(V)\det[J,\Theta]_{\mathrm{PB}},
\label{eq:8}
\end{eqnarray}
where $\Pi_i$ are conjugates of the angular variables $\phi_i$ defined by the 
linear transformation 
\begin{eqnarray}
\Pi_i&=&\sum_jV_{ij}L_j,\\
V&=&\left(\begin{array}{ccc}
-\sin\phi_2\cos\phi_3&\sin\phi_3&0\\
 \sin\phi_2\sin\phi_3&\cos\phi_3&0\\
 \cos\phi_2          &0         &1
\end{array}\right).
\end{eqnarray}
Let us now define $Z_{I}$ by imposing on Eq.\ (\ref{eq:8}) the boundary 
condition $L_i\rightarrow I_i$ for $\tau\rightarrow 0$ or $\beta$, where the 
total angular momentum $I$ is given by $I=\sqrt{\sum_iI_i^2}$\@. Then the path 
integral over $\phi_i$ and $L_i$ can be computed by standard methods: 
\begin{eqnarray}
Z_I&=&\int d^3I\int\prod_\alpha{\cal D}[b^\dagger_\alpha]{\cal D}[b_\alpha]
e^{-S_E(\beta)}\delta\left(I-\sqrt{\sum_iI_i^2}\right)\nonumber\\
&&\times\delta(\hbar I_i-J_i(b^\dagger,b))\delta(\Theta_i(b^\dagger,b))\det[J,
\Theta]_{\mathrm{PB}}.
\label{eq:11}
\end{eqnarray}
It is now useful to specify the $\Theta_i(b^\dagger,b)$ as the conjugate 
angular variables of the $J_i(b^\dagger,b)$\@. Then, the gauge condition 
$\delta(\Theta_i(b^\dagger,b))$ fixes the position of the intrinsic frame of 
reference relative to the rotating body. This eliminates rotations of the 
system as degrees of freedom associated with the fermion fields. However, the 
components $I_i$ of angular momentum in the intrinsic frame depend upon the 
choice of the gauge angular variables $\Theta_i(b^\dagger,b)$. 

\section{Model}
\label{sec3}

\subsection{The P+QQ Hamiltonian}

Let us consider the P+QQ Hamiltonian \cite{Kisslinger,Sorensen} as a model 
Hamiltonian 
\begin{eqnarray}
H&=&H_{\mathrm{s.p.}}-GP^\dagger P-\frac{\chi}{2}\sum_\mu Q_{2\mu}^\dagger
Q_{2\mu},\\
H_{\mathrm{s.p.}}&=&\sum_\alpha\varepsilon_\alpha c_\alpha^\dagger c_\alpha,\\
P&=&\sum_\alpha c_{\bar{\alpha}} c_\alpha,\\
Q_{2\mu}&=&\sum_\alpha\langle\alpha|r^2Y_{2\mu}|\beta\rangle c_\alpha^\dagger
c_\beta,
\end{eqnarray}
where $H_{\mathrm{s.p.}}$ is a single-particle Hamiltonian, $P$ is the monopole
pairing operator, $G$ is the strength of the pairing interaction, $Q_{2\mu}$ 
are components of the mass quadrupole tensor, and $\chi$ is the strength of the
quadru\-pole-quadrupole interaction. This model has been applied to the SPA+RPA
method including thermal and quantal fluctuations \cite{Puddu2}\@. 

\subsection{The Hubbard-Stratonovich transformation}

Using the Hubbard-Stratonovich transformation \cite{Hubbard} 
\begin{eqnarray}
\lefteqn{{\mathrm{const}}=\int{\cal D}\zeta^\ast{\cal D}\zeta\prod_\mu{\cal D}
[\sigma_\mu^\ast]{\cal D}[\sigma_\mu]}\nonumber\\
&&\times\exp\left[{\int_0^\beta d\tau G(\zeta^\ast-P^\dagger)(\zeta-P)}\right]
\nonumber\\
&&\times\exp\left[{\int_0^\beta d\tau\frac{\chi}{2}\sum_\mu(\sigma_\mu^\ast-
Q_{2\mu}^\dagger)(\sigma_\mu-Q_{2\mu})}\right],
\end{eqnarray}
the partition function of Eq.\ (\ref{eq:11}) can be written as the auxiliary 
field path integral \cite{Rossignoli,Lang} 
\begin{eqnarray}
\lefteqn{Z_I=\int d^3I\int{\cal D}\zeta^\ast{\cal D}\zeta\int\prod_\mu{\cal D}
[\sigma_\mu^\ast]{\cal D}[\sigma_\mu]{\mathrm{Tr}}\left[e^{-\beta H^\prime}
\right]}\nonumber\\
&&\times\exp\left[\int_0^\beta d\tau\left(-\frac{|\zeta|^2}{G}-\frac{\chi}{2}
|\sigma_\mu|^2\right)\right]\delta\left(I-\sqrt{\sum_iI_i^2}\right)\nonumber\\
&&\times\delta(\hbar I_i-J_i(b^\dagger,b))\delta(\Theta_i(b^\dagger,b))
\det[J,\Theta]_{\mathrm{PB}},
\label{eq:17}
\end{eqnarray}
where 
\begin{eqnarray}
{\mathrm{Tr}}\left[e^{-\beta H^\prime}\right]&=&\int\prod_\alpha{\cal D}
[b_\alpha^\dagger]{\cal D}[b_\alpha]\exp\left[-\int_0^\beta d\tau H^\prime
\right],\hspace*{0.7cm}\\
H^\prime&=&H_{\mathrm{s.p.}}-\zeta P^\dagger-\zeta^\ast P-\chi\sum_\mu
\sigma_\mu Q_{2\mu}^\dagger.
\label{eq:19}
\end{eqnarray}

\subsection{The SPA+RPA method}

In the perturbative expansion of the partition function $Z_I$ around the static
fields, the zeroth-order term gives the Hartree-like solution, while the 
second-order terms lead to the RPA corrections. This approximation may be 
largely improved by applying it to all time-independent paths of $\zeta$ and 
$\sigma_\mu$\@. We therefore expand the partition function (\ref{eq:17}) around
the static paths $\bar{\zeta}$ and $\bar{\sigma}_\mu$ in Fourier series, 
\begin{eqnarray}
\zeta&=&\bar{\zeta}+\sum_{n\ne 0}\xi_ne^{-i\omega_n\tau},\\
\sigma_\mu&=&\bar{\sigma}_\mu+\sum_{n\ne 0}\eta_{\mu n}e^{-i\omega_n\tau},
\end{eqnarray}
where $\omega_n=2\pi n/\beta$ are the Matsubara frequencies. Let us now define 
the principal axis of the quadrupole potential in the intrinsic frame by 
$\bar{\sigma}_0=\hbar\omega_0\beta_2\cos\gamma$, 
$\bar{\sigma}_2=\bar{\sigma}_{-2}=\hbar\omega_0\beta_2\sin\gamma/\sqrt{2}$, and
$\bar{\sigma}_1=\bar{\sigma}_{-1}=0$, where 
$\hbar\omega_0=41\,{\mathrm{MeV}}\,/A^{1/3}$\@. The pairing fields can be 
rewritten as $\bar{\zeta}=\Delta e^{-\psi}$ using real values $\Delta$ and 
$\psi$\@. 

It is now convenient to introduce quasiparticles by diagonalizing the 
Hamiltonian of Eq.\ (\ref{eq:19}) 
\begin{equation}
\left(\begin{array}{c}a\\a^\dagger\end{array}\right)={\cal W}^\dagger
\left(\begin{array}{c}b\\b^\dagger\end{array}\right)=
\left(\begin{array}{cc}\bar{U}^\dagger&\bar{V}^\dagger\\
\bar{V}^T&\bar{U}^T\end{array}\right)
\left(\begin{array}{c}b\\b^\dagger\end{array}\right),
\end{equation}
where the matrix ${\cal W}$ satisfies the unitarity condition 
${\cal W^\dagger}{\cal W}=1$\@. According to the Bloch-Messiah theorem 
\cite{Bloch}, the above unitary matrix can be decomposed into three matrices 
\begin{equation}
{\cal W}=\left(\begin{array}{cc}D&0\\0&D^\ast\end{array}\right)
\left(\begin{array}{cc}U&V\\V&U\end{array}\right)
\left(\begin{array}{cc}C&0\\0&C^\ast\end{array}\right).
\label{eq:23}
\end{equation}
The first transformation $D$ is determined by diagonalizing the deformed term 
of Eq.\ (\ref{eq:19}) 
\begin{eqnarray}
H_{\mathrm{def}}&=&H_{\mathrm{s.p.}}-\chi\sum_\mu\bar{\sigma}_\mu
Q_{2\mu}^\dagger\nonumber\\
&=&H_{\mathrm{s.p.}}-\hbar\omega_0\beta_2\nonumber\\
&&\times\left(Q_{20}\cos\gamma+\frac{Q_{22}+Q_{2-2}}{\sqrt{2}}\sin\gamma
\right),
\end{eqnarray}
from which we obtain the deformed fermion fields 
\begin{equation}
d_k=\sum_\alpha D_{k,\alpha}b_\alpha,
\end{equation}
where $D_{k,\alpha}$ are the matrix elements obtained by solving the eigenvalue
problem $H_{\mathrm{def}}|k\rangle=\varepsilon_k|k\rangle$\@. Now, the 
Hamiltonian $H^\prime$ of Eq.\ (\ref{eq:19}) can be simplified to 
\begin{equation}
H^\prime=\sum_k\varepsilon_kd_k^\dagger d_k-\Delta(P^\dagger+P).
\label{eq:26}
\end{equation}
The matrices $U$ and $V$ of Eq.\ (\ref{eq:23}) diagonalize the pairing term 
of Eq.\ (\ref{eq:26})\@. The matrices are diagonal and determined by solving 
the Hartree-Fock-Bogolyubov equations 
\begin{equation}
\left(\begin{array}{cc}\bar{\varepsilon}_k&\Delta\\\Delta&-\bar{\varepsilon}_k
\end{array}\right)
\left(\begin{array}{c}u_k\\v_k\end{array}\right)
=E_k\left(\begin{array}{c}u_k\\v_k\end{array}\right),
\end{equation}
where $E_k$ are the quasiparticle energies. 

In a small system such as the nucleus in which the particle number is strictly 
fixed, the canonical partition function should be employed. However, exact 
number projection is difficult to implement in the SPA+RPA formalism; it is 
easier to apply number parity projection. Performing the Gaussian integral over
$\xi_n$ and $\eta_{\mu n}$ in second order perturbation theory and introducing 
the number parity projection $P_s=(1+se^{i\pi N})/2$ where $s$ denotes the even
or odd number parity \cite{Rossignoli}, we obtain the partition function 
\begin{eqnarray}
\lefteqn{Z_{s,I}=\frac{2\beta}{G}\left(\frac{\kappa\beta}{2\pi}\right)^{5/2}}
\nonumber\\
&&\times\int_0^\infty d\Delta\Delta\int_0^\infty d\beta_2\beta_2^4
\int_0^{\pi/3}d\gamma|\sin(3\gamma)|\nonumber\\
&&\times 4\pi I^2\exp\left[-\left(\frac{\Delta^2}{G}+\frac{1}{2}\kappa\beta_2^2
\right)\beta\right]{\mathrm{Tr}}\left[P_se^{-\beta H^\prime}\right]\nonumber\\
&&\times\delta(\hbar I_i-J_i(a^\dagger,a))\delta(\Theta_i(a^\dagger,a))\det[J,
\Theta]_{\mathrm{PB}}C_{\mathrm{RPA}},\hspace*{0.7cm}
\label{eq:28}
\end{eqnarray}
where the parameter $\kappa=(\hbar\omega_0/b^2)^2/\chi$ and $b\propto A^{1/3}$ 
is the harmonic oscillator length. The expression (\ref{eq:28}) contains both 
the static mean-field contributions and the associated Gaussian corrections. 
These small-amplitude quantal corrections are included in the factor 
\begin{equation}
C_{\mathrm{RPA}}=C_{\mathrm{RPA}}(Q)C_{\mathrm{RPA}}(\Delta),
\end{equation}
where $C_{\mathrm{RPA}}(Q)$ and $C_{\mathrm{RPA}}(\Delta)$ are the corrections 
due to the $QQ$ and pairing correlations, respectively. The SPA partition 
function is obtained by neglecting these RPA corrections. $C_{\mathrm{RPA}}(Q)$
is explicitly given by 
\begin{equation}
C_{\mathrm{RPA}}(Q)=\prod_{n>0}\det(1-\chi R(\omega_n))^{-1}.
\end{equation}
Here $R(\omega_n)$ is the response function matrix given by 
\begin{eqnarray}
R(\omega_n)&=&\sum_{kl}\frac{Q_{kl}Q_{kl}(E_k+E_l)(1-f_k-f_l)}{(E_k+E_l)^2+
\omega_n^2}\nonumber\\
&&+\sum_{kl}\frac{\tilde{Q}_{kl}\tilde{Q}_{kl}(E_k-E_l)(f_k-f_l)}{(E_k-E_l)^2+
\omega_n^2},\\
Q_{kl}&=&q_{kl}(u_kv_l+v_ku_l),\\
\tilde{Q}_{kl}&=&q_{kl}(u_ku_l-v_kv_l),
\end{eqnarray}
where $q_{kl}$ are quadrupole matrix elements and $f_k$ are Fermi occupation 
probabilities $f_k=(1+e^{\beta E_k})^{-1}$\@. 
The response function can be calculated by solving the 
dispersion equation $\det(1-\chi R(i\omega_n))=0$\@. It can be also obtained by
diagonalizing the finite-temperature RPA equations \cite{Tanabe,Civitarese} 
\begin{equation}
\langle[H_{\mathrm{RPA}}(Q),O_\nu^\dagger]\rangle=\hbar\Omega_\nu\langle
O_\nu^\dagger\rangle,
\label{eq:34}
\end{equation}
where $\Omega_\nu$ are the RPA frequencies and the RPA modes are defined by 
\begin{eqnarray}
O_\nu^\dagger&=&\sum_{kl}(X_{kl}^\nu a_k^\dagger a_l^\dagger-Y_{kl}^\nu a_la_k)
\nonumber\\
&&+\sum_{kl}(Z_{kl}^\nu a_k^\dagger a_l-\bar{Z}_{kl}^\nu a_l^\dagger a_k).
\end{eqnarray}
The symbol $\langle\ \rangle$ in Eq.\ (\ref{eq:34}) denotes thermal averaging 
with respect to quasiparticles according to Wick's theorem 
\begin{eqnarray}
\langle F\rangle&=&\left.{\mathrm{Tr}}\left[F\exp\left(-\beta\sum_kE_k
a_k^\dagger a_k\right)\right]\right/Z_{\mathrm{qp}},\\
Z_{\mathrm{qp}}&=&{\mathrm{Tr}}\left[\exp\left(-\beta\sum_kE_ka_k^\dagger a_k
\right)\right],
\end{eqnarray}
where, as an an example, we give 
$\langle a_k^\dagger a_l\rangle=\delta_{kl}f_k$\@. At this point, it would be 
more correct to use thermal average and Fermi occupation probabilities in number 
parity projection instead, but the corresponding corrections in the RPA are 
normally small, and would not change our conclusions. It would be important 
for the odd-even effects in the moment of inertia \cite{Alhassid1}. The RPA 
Hamiltonian $H_{\mathrm{RPA}}(Q)$ can now be expressed in a diagonal form 
\begin{equation}
H_{\mathrm{RPA}}(Q)=\sum_\nu\hbar\Omega_\nu O_\nu^\dagger O_\nu.
\end{equation}
There are, however, zero-frequency modes $O_{\nu=0}$ due to the symmetry 
breaking of the deformed Hartree solutions on the static path. The correct 
treatment of these modes will be given in the next section. The RPA correction 
$C_{\mathrm{RPA}}(Q)$ can now be written exactly in terms of the RPA 
frequencies $\Omega_\nu$ 
\begin{equation}
C_{\mathrm{RPA}}(Q)=\frac{\prod_{kl}^\prime\frac{1}{(E_k+E_l)}\sinh\frac{\beta
(E_k+E_l)}{2}}{\prod_\nu\frac{1}{\Omega_\nu}\sinh\frac{\beta\Omega_\nu}{2}},
\end{equation}
where the prime in $\prod_{kl}^\prime$ denotes the restriction of the product 
to pairs $(k,l)$ that satisfy the conditions $k<l$ and $E_k+E_l\neq 0$\@. Note 
that for deformed or heavy nuclei there are many RPA frequencies $\Omega_\nu$ 
in the numerator and many $E_k+E_l$ pairs in the denominator. 

The pairing RPA correction $C_{\mathrm{RPA}}(\Delta)$ is obtained in a similar 
way as $C_{\mathrm{RPA}}(Q)$ by 
\begin{equation}
C_{\mathrm{RPA}}(\Delta)=\prod_k\frac{\omega_k\sinh[\beta E_k]}{2E_k\sinh[\beta
\omega_k/2]}.
\label{eq:40}
\end{equation}
Here, $\omega_k$ are the conventional thermal RPA energies and 
$E_k=\sqrt{\varepsilon_k^{\prime 2}+\Delta^2}$ with 
$\varepsilon_k^\prime=\varepsilon_k-\mu-G/2$\@. 

\section{Rotational motion and zero-frequency modes}
\label{sec4}

We consider the two-body Hamiltonian $H$ which is invariant under a continuous 
rotational operation generated by the angular momentum operators $J_{i}$\@. 
However, the deformed solutions on the static path violate this symmetry and 
lead to zero-frequency modes which are regarded as spurious modes in the 
intrinsic frame. Then, the rotational invariance must be restored by the 
residual interaction, which is defined as the difference between the exact and 
the mean-field Hamiltonian. Proper inclusion of the residual interaction will 
therefore restore the rotational invariance and will provide the rotational 
energy. To achieve this goal, we present in this section the correct treatment 
of zero-frequency modes within the finite temperature RPA\@. It is shown that 
the zero-frequency modes in the intrinsic frame are not spurious in the 
laboratory frame and correspond to three-dimensional rotational motions. 

\subsection{Treatment of zero-frequency modes}

For the calculation of the partition function of the constraint system (see 
Eq.\ (\ref{eq:28})), we follow the treatment proposed by Marshalek and Weneser 
\cite{Marshalek}\@. For the sake of simplicity, we assume an axially symmetric 
mean field for the deformed Hartree calculation with one violated symmetry and 
neglect the $\gamma$ degree of freedom, for simplicity. In particular, we 
consider the case where the symmetry is violated for rotations around the $x$ 
axis which is perpendicular to the symmetry or $z$ axis. Three-dimensional 
rotation can be treated in a similar way. 

When the eigenvalue equation (\ref{eq:34}) has zero-frequency solutions due to 
the breaking of rotational invariance in the intrinsic frame,\footnote{Note 
that the zero-frequency mode (Goldstone mode) in the mean field plus RPA may 
become non-zero (imaginary or complex) in the SPA+RPA\@.} the usual treatment 
is known to cause problems concerning the completeness and normalization. 
Therefore, instead of $O_\nu$ and $O_\nu^\dagger$, we define the following 
coordinate and conjugate momenta 
\begin{eqnarray}
p_\nu&=&\sqrt{\frac{\hbar\Omega_\nu}{2}}(O_\nu+O_\nu^\dagger),\\
q_\nu&=&i\sqrt{\frac{\hbar}{2\Omega_\nu}}(O_\nu-O_\nu^\dagger).
\end{eqnarray}
The finite-temperature RPA equations are then expressed as 
\begin{eqnarray}
\langle[H_{\mathrm{RPA}},q_\nu]_{\mathrm{PB}}\rangle&=&-i\hbar\langle p_\nu
\rangle,\\
\langle[H_{\mathrm{RPA}},p_\nu]_{\mathrm{PB}}\rangle&=&i\hbar\Omega_\nu^2
\langle q_\nu\rangle.
\end{eqnarray}
For the zero-frequency mode $O_{\nu=0}$ in the finite-tempera\-ture RPA 
equations (\ref{eq:34}), the angular momentum $J_x$ and the canonical conjugate
angle coordinate $\Theta_x$ are defined by $J_x={\cal J}_x^{1/2}p_0$ and 
$\Theta_x={\cal J}_x^{-1/2}q_0$, respectively, where ${\cal J}_x$ is the moment
of inertia. $\Theta_x$ and ${\cal J}_x$ can be determined by solving the 
equations of motion 
\begin{eqnarray}
\langle[H_{\mathrm{RPA}},\Theta_x]_{\mathrm{PB}}\rangle&=&\frac{-i\hbar}
{{\cal J}_x}\langle J_x\rangle,
\label{eq:45}\\
\langle[\Theta_x,J_x]_{\mathrm{PB}}\rangle&=&i\hbar,
\label{eq:46}
\end{eqnarray}
where Eq.\ \ref{eq:45} simply corresponds to the self-consistency condition 
$\dot{\Theta}_x=[H_{\mathrm{RPA}}-J_x^2/2{\cal J}_x,\Theta_x]_{\mathrm{PB}}=0$,
and we impose the gauge fixing condition $\Theta_x$=0 of Eq.\ (\ref{eq:28})\@. 
$H_{\mathrm{RPA}}(Q)$ can now be separated into an intrinsic and rotational 
part 
\begin{equation}
H_{\mathrm{RPA}}(Q)=\sum_{\nu>0}\hbar\Omega_\nu O_\nu^\dagger O_\nu+
\frac{J_x^2}{2{\cal J}_x}.
\end{equation}
Thus, the zero-frequency mode can be eliminated from the RPA equation and 
$C_{\mathrm{RPA}}(Q)$ can be written as 
\begin{equation}
C_{\mathrm{RPA}}(Q)=C_{\mathrm{RPA}}^\prime(Q)e^{-\frac{J_x^2}{2{\cal J}_x}
\beta},
\label{eq:48}
\end{equation}
where $C_{\mathrm{RPA}}^\prime(Q)$ is the correlation factor in the absence of 
zero-frequency modes. 

In a similar fashion, spurious modes should be eliminated from the pairing RPA 
corrections of Eq.\ (\ref{eq:40})\@. This is not done in the present work since
(i) exact number projection is needed for this procedure and (ii) the 
corrections are thought to be small in our number parity projection scheme.

\subsection{Moment of inertia}

In order to connect with the result of Thouless and Valatin \cite{Thouless}, we
introduce the function $G$ 
\begin{eqnarray}
G=i{\cal J}_x\Theta_x/\hbar&=& \sum_{k,l}\left[g_{kl}(u_kv_la_k^\dagger
a_{\bar{l}}^\dagger-v_ku_la_{\bar{l}} a_k)\right.\nonumber\\
&&-\left.\tilde{g}_{kl}(u_ku_la_k^\dagger a_l-v_kv_la_{\bar{l}}^\dagger
a_{\bar{k}})\right].
\label{eq:49}
\end{eqnarray}
Then the Thouless-Valatin equations become 
\begin{eqnarray}
\langle[H_{\mathrm{RPA}},G]_{\mathrm{PB}}\rangle&=&\langle J_x\rangle,\\
\langle[J_x,G]_{\mathrm{PB}}\rangle&=&{\cal J}_x.
\end{eqnarray}
Inserting Eq.\ (\ref{eq:49}) in the Thouless-Valatin equations yields 
\begin{eqnarray}
\hbar\left(j_x\right)_{kl}&=&(E_k+E_l)g_{kl}-\chi\sum_{mn}\left[Q_{kl}g_{mn}-
\tilde{Q}_{kl}\tilde{g}_{mn}\right.\nonumber\\
&&\hspace*{1.5cm}\left.+Q_{kl}g_{mn}^\ast-\tilde{Q}_{kl}\tilde{g}_{mn}^\ast
\right]Q_{mn},
\label{eq:52}\\
\hbar\left(j_x\right)_{kl}&=&(E_k-E_l)\tilde{g}_{kl}-\chi\sum_{mn}\left[Q_{mn}
g_{mn}-\tilde{Q}_{mn}\tilde{g}_{mn}\right.\nonumber\\
&&\hspace*{1.5cm}\left.+Q_{mn}g_{mn}^\ast-\tilde{Q}_{mn}\tilde{g}_{mn}^{\ast}
\right]\tilde{Q}_{kl},
\label{eq:53}
\end{eqnarray}
and the moment of inertia 
\begin{eqnarray}
{\cal J}_x&=&\hbar\sum_{k,l}(j_x)_{kl}\left[g_{lk}(u_kv_l-v_ku_l)^2(1-f_k-f_l)
\right.\nonumber\\
&&\hspace*{1cm}\left.+\tilde{g}_{lk}(u_ku_l+v_kv_l)^2(f_k-f_l)\right],
\end{eqnarray}
where we used the relations 
\begin{equation}
\begin{array}{rclcrcl}
g_{lk}&=&-g_{\bar{k}\bar{l}},&\hspace*{1cm}&\tilde{g}_{lk}&=&-
\tilde{g}_{\bar{k}\bar{l}},\\
u_k&=&u_{\bar{k}},&&v_k&=&-v_{\bar{k}}.
\end{array}
\end{equation}
Neglecting the two-body potential matrix elements in Eqs.\ (\ref{eq:52}) and 
(\ref{eq:53}), $g_{lk}$ and $\tilde{g}_{lk}$ can be approximated by
\begin{equation}
g_{kl}=\hbar\frac{(j_x)_{kl}}{(E_k+E_l)},\hspace*{1cm}
\tilde{g}_{kl}=\hbar\frac{(j_x)_{kl}}{(E_k-E_l)},
\end{equation}
and the moment of inertia at finite temperature simplifies to 
\begin{eqnarray}
\lefteqn{{\cal J}_x=2\hbar^2\sum_{k,l>0}\frac{(j_x)_{kl}^2}{(E_k+E_l)}(u_kv_l-
v_ku_l)^2(1-f_k-f_l)}\nonumber\\
&&+2\hbar^2\sum_{k,l>0}\frac{(j_x)_{kl}^2}{(E_k-E_l)}(u_ku_l+v_kv_l)^2(f_k-
f_l).
\end{eqnarray}
In the $T\rightarrow 0$ limit, the moment of inertia becomes the well-known 
Belyaev formula \cite{Belyaev} 
\begin{equation}
{\cal J}_x=2\hbar^2\sum_{k,l>0}\frac{(j_x)_{kl}^2}{(E_k+E_l)}(u_kv_l-v_ku_l)^2.
\end{equation}
Inserting Eqs.\ (\ref{eq:46})--(\ref{eq:48}) into the partition function 
(\ref{eq:28}) under consideration of the gauge fixing conditions $\hbar I=J_x$ 
and $\Theta_x=0$, we obtain the partition function for an axially symmetric 
nucleus 
\begin{eqnarray}
\lefteqn{\hspace*{-2cm}Z_{s,I}=\frac{2\beta}{G}\left(\frac{\kappa\beta}{2\pi}
\right)^{5/2}\int_0^\infty d\Delta\Delta\int_0^\infty d\beta_2\beta_2^4Z_s
(\Delta,\beta_2)}\nonumber\\
&&\times4\pi I^2\exp\left(-\frac{\hbar^2I^2}{2{\cal J}_x}\beta\right),
\label{eq:59}
\end{eqnarray}
where 
\begin{eqnarray}
Z_s(\Delta,\beta_2)&=&\exp\left[-\left(\frac{\Delta^2}{G}+\frac{1}{2}\kappa
\beta_2^2\right)\beta\right]\nonumber\\
&&\times Z_s^{\mathrm{qp}}C_{\mathrm{RPA}}^\prime(Q)C_{\mathrm{RPA}}(\Delta),
\label{eq:60}
\end{eqnarray}
and $Z_s^{\mathrm{qp}}$ is the quasiparticle partition function with the number 
parity projection
\begin{eqnarray}
Z_s^{\mathrm{qp}}&=&\frac{1}{2}\prod_ke^{-\gamma_k\beta}(1+e^{-E_k\beta})^2
\nonumber\\
&&\times\left[1+s\prod_{k^\prime}\tanh^2(E_{k^\prime}\beta)\right],
\end{eqnarray}
with $\gamma_k=\varepsilon_k-\mu-E_k$\@.

\subsection{Level density}

\begin{figure}[t!]
\includegraphics[totalheight=10.2cm]{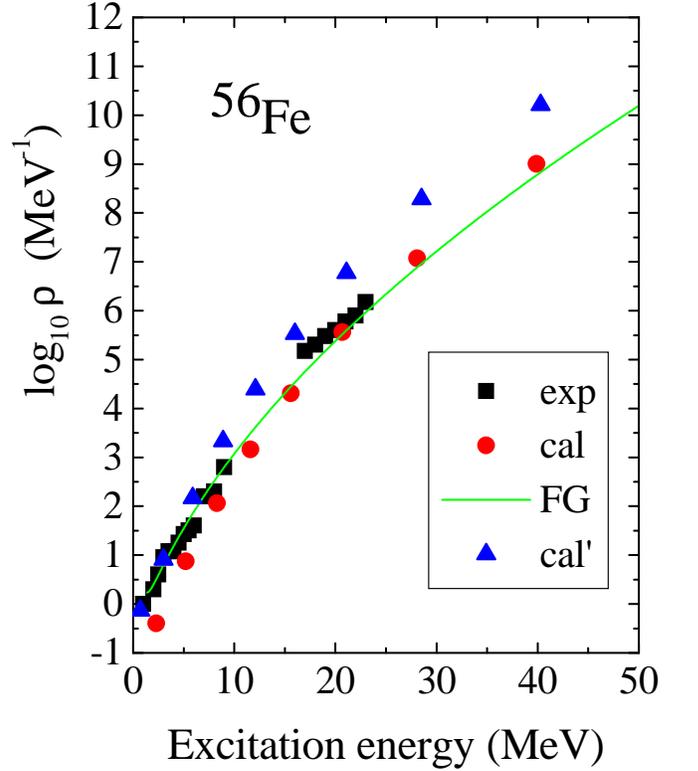}
\caption{(Color online) Experimental (squares) and calculated level densities 
$\rho(E)$ as function of excitation energy in $^{56}$Fe. Full circles and 
triangles denote the level and state density, respectively, from SPA+RPA 
calculations. The solid line is a back-shifted Fermi-gas model.}
\label{fig1}
\end{figure}

The level density for a system with a fixed number of particles can be 
evaluated by an inverse Laplace transformation of the partition function 
$Z_{s,I}$ 
\begin{equation}
\rho(E,I)=\frac{1}{2\pi i}\int_{\beta_0-i\infty}^{\beta_0+i\infty}d\beta
e^{\beta E}Z_{s,I}(\beta).
\end{equation}
In the saddle-point approximation, the level density is given by 
\begin{equation}
\rho(E,I)\approx\frac{Z_{s,I}e^{\beta E}}{[2\pi\partial^2\ln Z_{s,I}/\partial
\beta^2]^{1/2}}.
\label{eq:63}
\end{equation}
We will now apply the saddle-point approximation\footnote{The maximum in the 
saddle-point approximation leads to an effective mean-field equation, hence, a 
sharp phase transition as predicted by the ordinary mean-field equation is 
avoided.} to the two integrals in the partition function $Z_{s,I}$ of Eq.\ 
(\ref{eq:59})\@. However, Eq.\ (\ref{eq:59}) still contains the $2I+1$ 
degeneracy connected to the magnetic quantum number. In order to determine the 
level density (as opposed to the state density), we make the transformation 
$I\rightarrow I+1/2$ and $I^2\rightarrow I(I+1)$ in Eq.\ (\ref{eq:59}) from a 
classical to a quantal angular momentum and divide by $2I+1$\@.\footnote{In 
order to keep in contact with experiment, we will consider this pseudo 
partition function for the remainder of this work and determine consistently 
all quantities from there.} Entering the thusly modified partition function
(\ref{eq:59}) into the saddle-point approximation for the level density 
(\ref{eq:63}), we finally obtain 
\begin{equation}
\rho(E,I)\approx\rho(E)\frac{\hbar^2\beta}{2{\cal J}_x}(2I+1)
\exp\left(-\frac{\hbar^2I(I+1)}{2{\cal J}_x}\beta\right),
\label{eq:64}
\end{equation}
where the total level density is given by 
\begin{equation}
\rho(E)=\frac{\bar{Z}_se^{\beta E}}{[2\pi\partial^2\ln\bar{Z}_s/\partial
\beta^2]^{1/2}},
\end{equation}
with 
\begin{equation}
\bar{Z}_s=\frac{2\beta}{G}\left(\frac{\kappa\beta}{2\pi}\right)^{5/2}
\frac{\Delta\beta_2^4Z_s(\Delta,\beta_2)}{\sqrt{\det(B)}},
\label{eq:66}
\end{equation}
and $B$ is the fluctuation matrix 
\begin{equation}
B_{ij}=\frac{\partial^2\ln\left(\Delta\beta_2^4Z_s(\Delta,\beta_2)\right)}
{\partial y_i\partial y_j},
\label{eq:67}
\end{equation}
with $y=(\Delta,\beta_2)$\@. 

\begin{figure}[t!]
\includegraphics[totalheight=9.8cm]{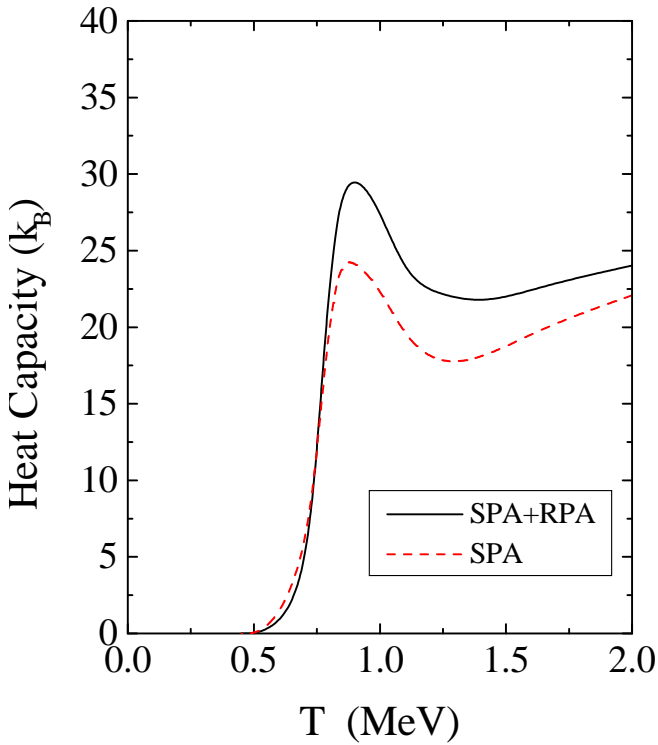}
\caption{(Color online) Heat capacities as a function of temperature $T$ in 
$^{56}$Fe. The solid and dotted curves denote the calculated heat capacities in
the SPA+RPA and SPA, respectively. Both curves exhibit clear $\cal{S}$ shapes.}
\label{fig2}
\end{figure}

The level density (\ref{eq:64}) is the relevant spin-dependent microscopic 
level density for reaction cross-section calculations in the presence of 
correlations. By entering the partition function (\ref{eq:60}) in the Eqs.\ 
(\ref{eq:66}) and (\ref{eq:67}), we can calculate this level density as a 
function of excitation energy, where the thermal energy can be calculated from 
$E=-\partial\ln\bar{Z}_s/\partial\beta$\@. Comparing this level density with 
the spin-cutoff model given in Eq.\ (\ref{eq:1}), the spin-cutoff parameter 
$\sigma^2$ can be related to the moment of inertia ${\cal J}_x$ by 
\begin{equation}
\sigma^2=\frac{{\cal J}_xT}{\hbar^2}.
\label{eq:68}
\end{equation}

\begin{figure}[t!]
\includegraphics[totalheight=10.7cm]{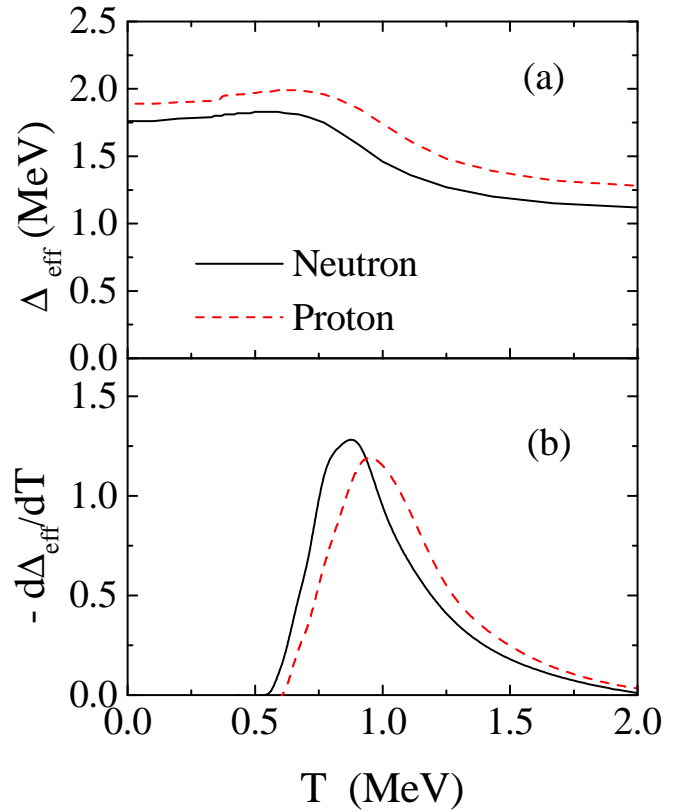}
\caption{(Color online) Effective pairing gaps (a) and their derivatives (b) as
a function of temperature for $^{56}$Fe. The solid and dotted curves are for 
neutrons and protons, respectively.}
\label{fig3}
\end{figure}

In the next section, we shall perform numerical calculations and show that 
there is a notable difference between the shape of the level density and the 
state density which contains the factor $2I+1$ due to the $m$-degeneracy. 

\section{Numerical calculations}
\label{sec5}

We consider $^{56}$Fe \cite{Huizenga,Lu} as an illustrative example. Recently, 
the level density in $^{56}$Fe has been measured nearly up to the neutron 
binding energy \cite{Schiller2,Tavukcu,Voinov}\@. In our calculation, we use 
the single-particle energies $\varepsilon_k$ given by an axially deformed 
Woods-Saxon potential with spin-orbit interaction \cite{Cwoik}\@. The 
Woods-Saxon parameters are chosen such as to approximately reproduce the 
experimental single-particle energies extracted from the energy levels of the 
odd nucleus $^{41}$Ca (a $^{40}$Ca core plus one neutron)\@. 15 
doubly-degenerate single-particle levels for neutrons and protons outside the 
$^{40}$Ca core are considered; continuum levels are neglected since their 
contributions are small \cite{Alhassid1}\@. We adjust the pairing-force 
strengths at $G_n=25/A$~MeV and $G_p=29/A$~MeV for neutrons and protons, 
respectively, and the QQ-force strength at $\chi=240/A^{5/3}b^4$\@. As 
mentioned above, the SPA+RPA breaks down at low temperature. However, it has 
recently been shown that in the monopole pairing case the SPA+RPA with 
number-parity projection reproduces well exact results even for low 
temperatures \cite{Rossignoli}\@. Therefore, number-parity projection is 
essential to describe the thermal properties at low temperature. Calculations 
are shown in Fig.\ \ref{fig1}. The level density calculated with the SPA+RPA 
reproduces fairly well the slope and magnitude of the experimental data. On the
other hand, beside the expected difference in magnitude, the calculated state 
density which contains the $m$-degeneracy shows also a different slope on the 
logarithmic plot. The calculations thus indicate an increase of 
$\sigma\propto\langle I\rangle$ with excitation energy, see also Eq.\ 
(\ref{eq:68})\@. 

\begin{figure}[t!]
\includegraphics[totalheight=10.8cm]{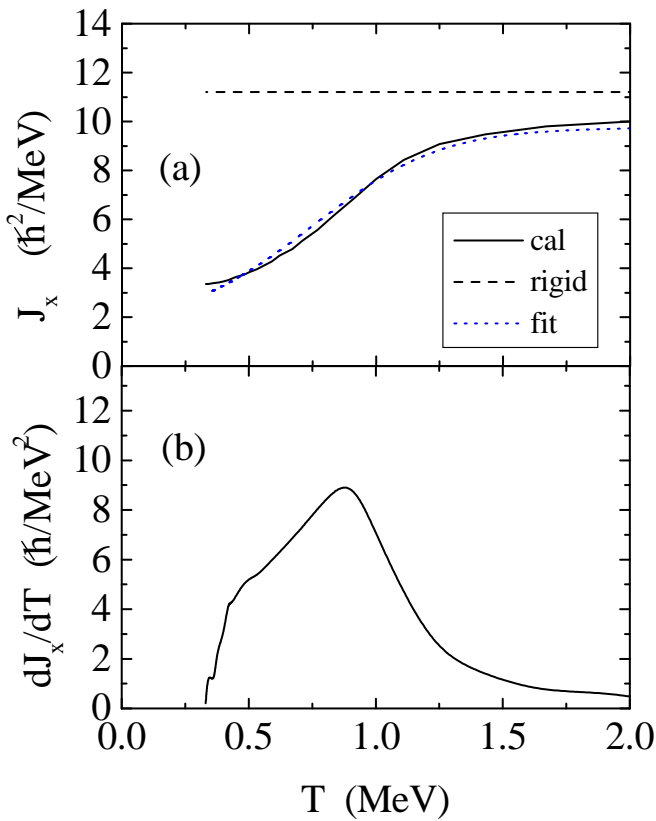}
\caption{(Color online) Moment of inertia (a) and the derivative (b) as 
function of temperature for $^{56}$Fe. The dashed line shows the rigid-body 
moment of inertia, the solid curve is the result of the SPA+RPA calculation, 
and the dotted curve is a fit to the calculation according to Eq.\ 
(\protect\ref{eq:73})\@.}
\label{fig4}
\end{figure}

Figure \ref{fig1} also shows a back-shifted Fermi-gas model 
\cite{Bethe,Gilbert,Grimes,Lu,Egidy} according to 
\begin{equation}
\rho_{\mathrm{BSFG}}(E)=\frac{\exp\left[2\sqrt{aU}\right]}{12\sqrt{2}a^{1/4}
U^{5/4}\sigma},
\end{equation}
where the back-shifted energy is $U=E-E_1$ and the spin-cutoff parameter 
$\sigma$ is $\sigma^2=0.0888A^{2/3}\sqrt{aU}$ \cite{Gilbert}\@. The 
level-density parameter $a$ and the parameter $E_1$ are given by 
$a=A/10=5.6$~MeV$^{-1}$ and $E_1=1.0$~MeV, respectively. The level density from
SPA+RPA calculations is close to the back-shifted Fermi-gas model for a wide 
range of excitation energies. 

\begin{figure}[t!]
\includegraphics[totalheight=7.1cm]{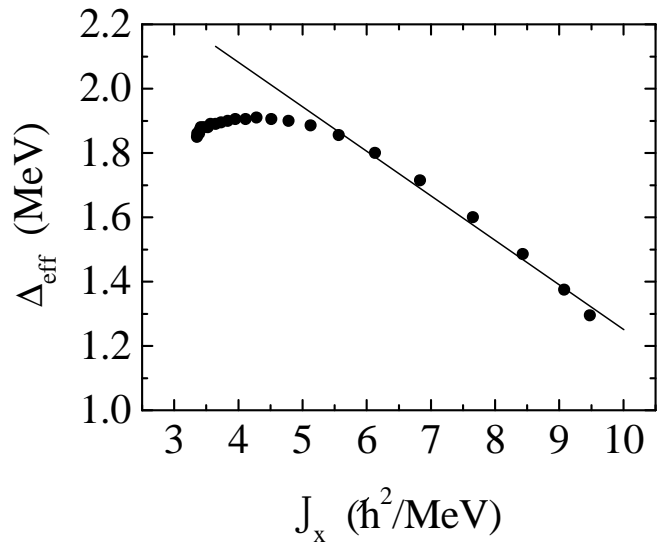}
\caption{Correlation of the average effective pairing gap with the moment of 
inertia. The straight line is to guide the eye.}
\label{fig5}
\end{figure}

Figure \ref{fig2}\@ shows the calculated heat capacity for $^{56}$Fe. The 
SPA+RPA deviates from the SPA result for temperatures above $T=$0.9~MeV; 
quantal fluctuations become important in this region. This behavior is 
consistent with the one found in Ref.\ \cite{Canosa}\@. Both of our results 
exhibit a characteristic $\cal{S}$-shape around $T_c\sim 0.9$~MeV\@. This 
critical temperature is higher than the SMMC result\footnote{We would like to 
emphasize that in contrast to the present calculation and the experiment, the 
SMMC result is obtained from a partition function which includes the $m$ 
degeneracy} $\sim 0.7$~MeV \cite{Liu} but it is still smaller than the 
experimental one $\sim 1.3$~MeV \cite{Tavukcu}\@. Level densities in several 
nuclei have recently been measured \cite{Bondarenko,Chankova} and the 
suppression of pairing correlations was studied in detail 
\cite{Kaneko1,Kaneko2}\@. $\cal{S}$-shaped heat capacities with 
$T_c\sim 0.5$~MeV were also observed in $^{162}$Dy, $^{166}$Er, and $^{172}$Yb 
\cite{Schiller,Melby}; they were interpreted as a signature of the breaking of 
nucleon Cooper pairs. We suggested that the suppression of pairing correlations
at finite temperatures should also appear in the thermal odd-even mass 
differences \cite{Kaneko1,Kaneko2}\@. In general, the $\cal{S}$-shape of the 
heat capacity has been attributed to the reduction of the pairing energy which 
can be calculated from 
$G_\tau\langle P^\dagger P\rangle=G_\tau T\partial\ln\bar{Z}_s/\partial G_\tau$
\cite{Liu}\@. In fact, this pairing reduction can be seen in thermal odd-even 
mass differences extracted from the experimental level densities 
\cite{Kaneko1,Kaneko2,Kaneko3}\@. On the other hand, it has been argued that 
the $\cal{S}$-shape might be accounted for as an effect of the particle-number 
conservation on the quasiparticle excitations \cite{Esashika}\@. However, even 
without exact particle-number projection, the SPA gives an $\cal{S}$-shaped 
heat capacity. In this sense, we emphasize that this $\cal{S}$ shape cannot be 
explained solely as an effect of the particle-number projection.  

Figure \ref{fig3}a shows the effective pairing gap defined by 
\begin{equation}
\Delta_{\mathrm{eff}}^\tau=G_\tau\left[\frac{1}{\beta}\frac{\partial\ln
\bar{Z}_s}{\partial G_\tau}\right]^{1/2}.
\end{equation}
$\Delta_{\mathrm{eff}}^n$ and $\Delta_{\mathrm{eff}}^p$ decrease in a similar 
fashion around $T=0.9$~MeV, although the proton pairing is a little bit 
stronger than the neutron pairing. We can now identify the inflection point of 
the $\Delta_{\mathrm{eff}}$ curves as the critical temperature of a pairing 
phase transition. In order to determine the position of the inflection point 
precisely, we differentiate the $\Delta_{\mathrm{eff}}$ curves with respect to 
the temperature, see Fig.\ \ref{fig3}b. From the peaks of these derivatives we 
read off a critical temperature around $T=0.9$~MeV\@. Hence, the suppression of
$\Delta_{\mathrm{eff}}$ is well correlated with the $\cal{S}$ shape of the heat
capacity. As pointed out in our previous paper \cite{Kaneko1}, the critical 
temperature $T_c$ is proportional to the pairing gap $\Delta$\@. This is 
expressed by the relation $T_c\approx0.57\Delta$ from Bardeen-Cooper-Schrieffer
(BCS) theory. The critical temperature $T_c\approx1.0$~MeV estimated from this 
relation is close to the position $\sim 0.9$~MeV of the peaks in Fig.\ 
\ref{fig3}b. 

\begin{figure}[t!]
\includegraphics[totalheight=10.8cm]{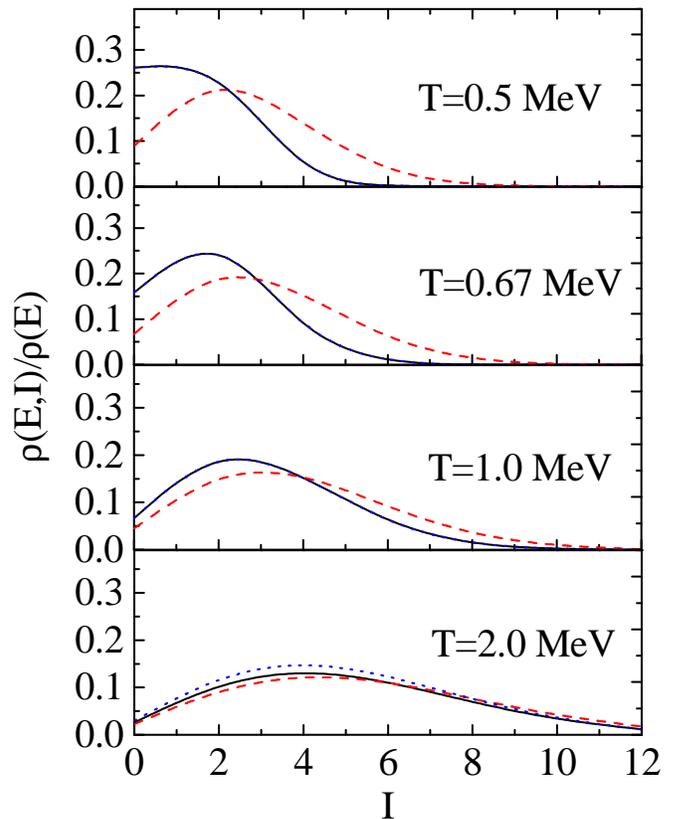}
\caption{(Color online) Spin distribution of level density at different 
temperatures $T$ for $^{56}$Fe. The SPA+RPA results (solid curves) are compared
to the spin-cutoff model using the rigid-body moment of inertia (dashed curves)
and the results from the fit to the moment of inertia according to Eq.\ 
(\protect\ref{eq:73}) (dotted curves)\@. The fit and the SPA+RPA calculations 
are almost indistinguishable.}
\label{fig6}
\end{figure}

We will now apply a similar analysis to the moment of inertia. Figure 
\ref{fig4}a shows the moment of inertia as function of temperature $T$\@. For 
comparison we also show the rigid-body value 
${\cal J}_{\mathrm{rigid}}=0.0137A^{5/3}$~MeV$^{-1}$ denoted by a dashed line. 
The calculated moment of inertia is smaller than the rigid-body value at low 
temperature. This strong suppression is well known as the effect due to the 
presence of pairing correlations in low-lying, superfluid-like BCS states. For 
high temperatures, however, the calculation approaches the rigid-body value 
because the nucleon Cooper pairs break with increasing temperature. Again, we 
can define the critical temperature of the pairing phase transition as the 
inflection point which is determined precisely by differentiation. 
Differentiation of the ${\cal J}_x$ curves reveals peaks around the critical 
temperature of $T=0.9$~MeV (see Fig.\ \ref{fig4}b)\@. Hence, the increase of 
${\cal J}_x$ is well correlated with both the $\cal{S}$ shape of the heat 
capacity in Fig.\ \ref{fig2} and the suppression of the effective pairing gaps 
in Fig.\ \ref{fig3}\@. 

\begin{figure}[t!]
\includegraphics[totalheight=9.1cm]{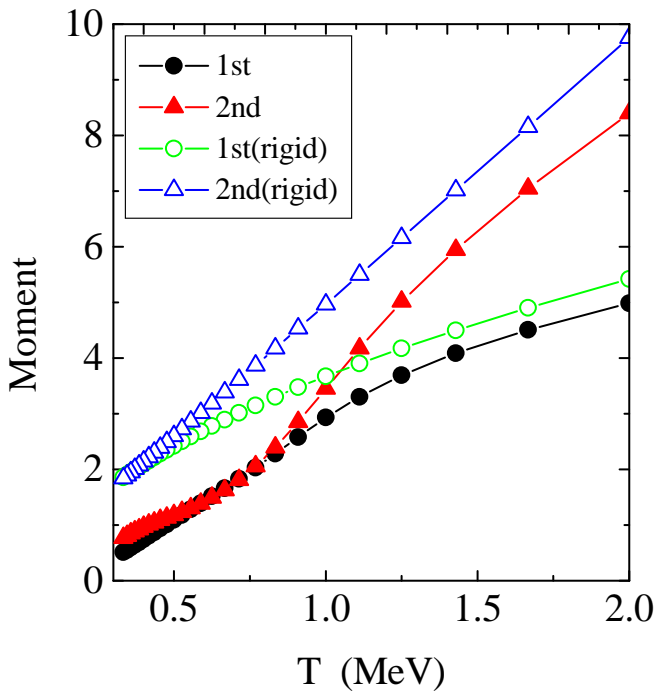}
\caption{(Color online) First (circles) and second (triangles) moments of the 
spin distribution as function of temperature for $^{56}$Fe. Results from the 
SPA+RPA calculation (full symbols) are compared to a model using the rigid-body
moment of inertia (open symbols)\@.}
\label{fig7}
\end{figure}

To investigate more closely the relation between the effective gaps and the 
moment of inertia, we examine the correlation of the average effective pairing 
gap $\Delta_{\mathrm{eff}}=(\Delta_{\mathrm{eff}}^n+\Delta_{\mathrm{eff}}^p)/2$
with the moment of inertia, see Fig.\ \ref{fig5}\@. Above 
${\cal J}_x\approx5.0$~$\hbar^2$/MeV, which corresponds to $T\approx0.7$~MeV, 
this correlation is close to a straight line, while deviations from the line 
become apparent below this temperature. The observation of such a correlation 
across the critical temperature of $T\approx0.9$~MeV means that the increase of
the moment of inertia can be attributed to the suppression of the pairing 
correlation. This conclusion has already been obtained by Alhassid {\it et 
al.\@} \cite{Alhassid1,Alhassid2}\@. Thus the increase of the moment of inertia
can be considered as another signature of the pairing phase transition together
with the $\cal{S}$-shape of the heat capacity \cite{Schiller,Melby} and the 
suppression of odd-even mass differences \cite{Kaneko1,Kaneko2,Kaneko3}\@. 

\begin{figure}[t!]
\includegraphics[totalheight=10.8cm]{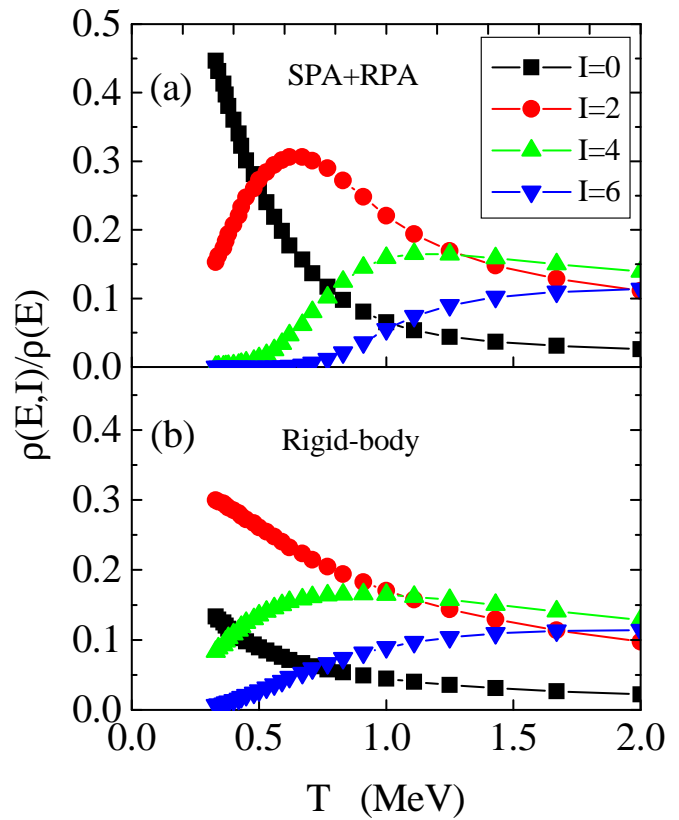}
\caption{(Color online) Spin components of level density as a function of 
temperature $T$ for $^{56}$Fe. Relative components for spins $J=0$, 2, 4, and 
6 from SPA+RPA calculations (upper panel) and from a spin-cutoff model using 
the rigid-body moment of inertia (lower panel)\@.}
\label{fig8}
\end{figure}

In the next step, we investigate the effect of the temperature-dependent moment
of inertia on the spin distribution of nuclear levels for the case $^{56}$Fe. 
For this purpose, we plot for the temperatures $T=0.5$, 0.67, 1.0, and 2.0~MeV 
the spin distribution $\rho(E,I)/\rho(E)$ from Eq.\ (\ref{eq:64}) in Fig.\ 
\ref{fig6}\@. Moreover, we plot as function of temperature the first $\mu_1$ 
and second $\mu_2$ moments of this spin distribution where 
\begin{eqnarray}
\mu_1&=&\frac{\sum_If_II}{\sum_II},\\
\mu_2&=&\frac{\sum_If_I(I-\mu_1)^2}{\sum_II},
\end{eqnarray}
and $f_I=\rho(E,I)/\rho(E)$ in Fig.\ \ref{fig7}, and finally, we plot again as 
function of temperature relative components of the spin distribution for the 
spins $I=0$, 2, 4, and 6~$\hbar$\@. Comparing the SPA+RPA results to the 
spin-cutoff model based on a rigid-body moment of inertia, we observe from 
Figs.\ \ref{fig6}, \ref{fig7}, and \ref{fig8} that good agreement between the 
two is achieved at high temperatures $T\agt2$~MeV\@. Deviations are, however, 
significant for lower temperatures. In particular, the $I=0$ component is much 
larger than all other $I$-components in the SPA+RPA calculation, whereas for 
the rigid-body spin-cutoff model, the largest component at low temperature is 
$I=2$ while the $I=0$ component is quite small. Around $T\sim 0.5$~MeV in the 
SPA+RPA calculation, the $I=0$ component quickly decreases with increasing 
temperature whereas the $I=2$ component increases drastically up to $T=0.7$~MeV
where it becomes dominant, in agreement with the rigid-body spin-cutoff model. 
The $I=4$ and 6 components also exhibit small magnitudes at low temperature and
increase with increasing temperature. 

To conclude the discussion, we would like to make two remarks. Firstly, the 
spin distributions in Fig.\ \ref{fig6} can be very well described by a 
phenomenological model for the moment of inertia 
\begin{equation}
\tilde{{\cal J}}_x={\cal J}_{\mathrm{rigid}}-\frac{c_1}{1+\exp(c_2T^2)}-c_3,
\label{eq:73}
\end{equation}
with the parameters $c_1=15.0$~$\hbar^2/$MeV, $c_2=1.8$~MeV$^{-2}$, and 
$c_3=1.5$~$\hbar^2/$MeV\@. This model of the moment of inertia includes the 
effect of the quenching of pairing correlations. The three parameters of the 
model essentially determine the moment of inertia at zero temperature (in the 
presence of pairing correlations), at very high temperature (in the absence of 
pairing correlations), and the critical temperature itself. Hence, the three 
parameters might be connected to collective $E2$ transition rates within 
rotational bands at low and high spin, respectively, and to odd-even mass 
differences via $T_c\approx0.57\Delta$\@. This would be interesting to 
investigate since a simple phenomenological model of the moment of inertia 
based on experimental observables would be valuable for many applications of 
the nuclear level density such as Hauser-Feshbach-type cross-section 
calculations. Secondly, we note the presence of an $\cal{S}$ shape in the 
second moment of the spin distribution (see Fig.\ \ref{fig7})\@. This is in 
analogy to the $\cal{S}$ shape of the heat capacity (see Fig.\ \ref{fig2}) 
which in itself is proportional to the second moment of the excitation-energy 
distribution for a given $T$\@. We would therefore propose this $\cal{S}$ shape
of the second moment as another signature of the pairing phase transition, and 
it would be interesting to look for experimental evidence for it. 

\section{Conclusion}
\label{sec6}

We have presented a treatment of nuclear rotation by the SPA+RPA method 
including thermal and quantal fluctuations. Using the collective coordinate 
method and applying a Marshalek-Weneser treatment, the RPA partition function 
is separated into two parts: an intrinsic RPA and a rotational part. We then 
can derive the spin-dependent level density $\rho(E,I)$ in the saddle-point 
approximation without the assumption of individual random spin vectors. The 
spin-cutoff parameter $\sigma$ can be identified with the moment of inertia 
${\cal J}_x$ of the generalized Belyaev formula. As an illustrative example, we
have applied this method to $^{56}$Fe and investigated the thermal properties 
including the moment of inertia with consideration of thermal and quantal 
fluctuations. We obtained the same conclusion as Alhassid {\it et al.\@} 
\cite{Alhassid1,Alhassid2}, namely that the increase of the moment of inertia 
with increasing temperature is attributed to the suppression of pairing 
correlations. The increase of the moment of inertia can therefore be considered
as one of the signatures of the pairing phase transition in parallel with the 
$\cal{S}$-shape of the heat capacity and the suppression of odd-even mass 
differences. The spin distribution of the level density shows that the $I=0$ 
component is dominant at low temperatures due to the presence of pairing 
correlation; this component decreases with increasing temperature. At high 
temperatures, the level density is governed by $I\ne0$ components. In this 
work, we assumed an axially symmetric nucleus with $K=0$ although our 
formulation can be easily extended to non-zero $K$ quantum numbers. However, 
the $K$ quantum number is thought to disappear due to $K$-mixing at some 
excitation energy above the yrast line. For this mixing, another degree of 
freedom, namely $\gamma$ deformation is important. It would be very interesting
to investigate this issue in the future. 

\acknowledgments

One of the authors (K.K.) would like to thank Dr.\ M. Hasegawa for fruitful and
inspiring discussions. Financial support from the National Science Foundation 
under grant number PHY-06-06007 is gratefully acknowledged.


\begin{thebibliography}{99}
\bibitem{Rauscher}T. Rauscher, F.-K. Thielemann, and K.-L. Kratz, Phys.\ Rev.\ 
C {\bf 56}, 1613 (1997).
\bibitem{Bethe}H.A. Bethe, Phys.\ Rev.\ {\bf 50}, 332 (1936).
\bibitem{Gilbert}A. Gilbert and A.G.W. Cameron, Can.\ J. Phys.\ {\bf 43}, 1446 
(1965).
\bibitem{Grimes}S.M. Grimes, J.D. Anderson, J.W. McClure, B.A. Pohl, and C. 
Wong, Phys.\ Rev.\ C {\bf 10}, 2373 (1974).
\bibitem{Lu}C.C. Lu, L.C. Vaz, and J.R. Huizenga, Nucl.\ Phys.\ {\bf A190}, 229
(1972).
\bibitem{Egidy}T. von Egidy, H.H. Schmidt, and A.N. Behkami, Nucl.\ Phys.\ 
{\bf A481}, 189 (1988); T. von Egidy and D. Bucurescu, Phys.\ Rev.\ C {\bf 
72}, 044311 (2005); {\bf 73}, 049901(E) (2006).
\bibitem{Ericson}T. Ericson, Adv.\ Phys.\ {\bf 9}, 425 (1960).
\bibitem{Alhassid1}Y. Alhassid, G.F. Bertsch, L. Fang, and S. Liu, Phys.\ Rev.\
C {\bf 72}, 064326 (2005).
\bibitem{Alhassid2}Y. Alhassid, S. Liu, and H. Nakada, nucl-th/0607062.
\bibitem{Alhassid3}Y. Alhassid and J. Zingman, Phys.\ Rev.\ C {\bf 30}, 684
(1984); Y. Alhassid and B.W. Bush, Nucl.\ Phys.\ {\bf A549}, 43 (1992).
\bibitem{Bertsch}P. Arve, G. Bertsch, B. Lauritzen, and G. Puddu, Ann.\ 
Phys.\ (N.Y.) {\bf 183}, 309 (1988); B. Lauritzen, P. Arve, and G.F. Bertsch, 
Phys.\ Rev.\ Lett.\ {\bf 61}, 2835 (1988).
\bibitem{Rossignoli1}R. Rossignoli, A. Ansari, and P. Ring, Phys.\ Rev.\ Lett.\
{\bf 70}, 1061 (1993); R. Rossignoli and P. Ring, Ann.\ Phys.\ (N.Y.) {\bf 
235}, 350 (1994).
\bibitem{Puddu1}G. Puddu, P.F. Bortignon, and R.A. Broglia, Ann.\ Phys.\ (N.Y.)
{\bf 206}, 409 (1991); Phys.\ Rev.\ C {\bf 42}, R1830 (1990).
\bibitem{Lauritzen}B. Lauritzen, G. Puddu, P.F. Bortignon, and R.A. Broglia, 
Phys.\ Lett.\ {\bf B246}, 329 (1990); B. Lauritzen, A. Anselmino, P.F. 
Bortignon, and R.A. Broglia, Ann.\ Phys.\ (N.Y.) {\bf 223}, 216 (1993). 
\bibitem{Rossignoli}R. Rossignoli, N. Canosa, and P. Ring, Phys.\ Rev.\ Lett.\ 
{\bf 80}, 1853 (1998); R. Rossignoli and N. Canosa, Phys.\ Lett.\ {\bf B394}, 
242(1997).
\bibitem{Attias}H. Attias and Y. Alhassid, Nucl.\ Phys.\ {\bf A625}, 565 
(1997).
\bibitem{Gervais1}J.-L. Gervais and B. Sakita, Phys.\ Rev.\ D {\bf 11}, 2943 
(1975); J.L. Gervais and A. Neveu, Phys.\ Rep.\ {\bf 23}, 237 (1976).
\bibitem{Gervais2}J.L. Gervais, A. Jevicki, and B. Sakita, Phys.\ Rep.\ {\bf 
23}, 281 (1976).
\bibitem{Alessandrini}V. Alessandrini, D.R. B\`{e}s, and B. Machet, Nucl.\ 
Phys.\ {\bf B142}, 489 (1978).
\bibitem{Bes}D.R. B\`{e}s, O. Civitarese, and H.M. Sof\'{\i}a, Nucl.\ Phys.\ 
{\bf A370}, 99 (1981).
\bibitem{Kisslinger}L.S. Kisslinger and R. A. Sorensen, Rev.\ Mod.\ Phys.\ {\bf
35}, 853 (1963).
\bibitem{Sorensen}D.R. Bes and R.A. Sorensen, Adv.\ Nucl.\ Phys.\ {\bf 2}, 129 
(1969).
\bibitem{Puddu2}G. Puddu, Phys.\ Rev.\ C {\bf 47}, 1067 (1993).
\bibitem{Hubbard}J. Hubbard, Phys.\ Rev.\ Lett.\ {\bf 3}, 77 (1959); R.L. 
Stra\-tonovich, Dokl.\ Akad.\ Nauk SSSR {\bf 115}, 1097 (1957).
\bibitem{Lang}G.H. Lang, C.W. Johnson, S.E. Koonin, and W.E. Ormand, Phys.\ 
Rev.\ C {\bf 48}, 1518 (1993); S.E. Koonin, D.J. Dean, and K. Langanke, Phys.\ 
Rep.\ {\bf 278}, 1 (1997).
\bibitem{Bloch}C. Bloch and A. Messiah, Nucl.\ Phys.\ {\bf 39}, 95 (1962).
\bibitem{Tanabe}K. Tanabe and N.D. Dang, Phys.\ Rev.\ C {\bf 62}, 024310 
(2000).
\bibitem{Civitarese}O. Civitarese, J.G. Hirsch, F. Montani, and M. Reboiro, 
Phys.\ Rev.\ C {\bf 62}, 054318 (2000).
\bibitem{Marshalek}E.R. Marshalek and J. Weneser, Ann.\ Phys.\ (N.Y.) {\bf 53},
569 (1969).
\bibitem{Thouless}D.J. Thouless and J.G. Valatin, Nucl.\ Phys.\ {\bf 31}, 211 
(1962).
\bibitem{Belyaev}S.T. Beliaev, Nucl.\ Phys.\ {\bf 24}, 322 (1961).
\bibitem{Huizenga}J.R. Huizenga, H.K. Vonach, A.A. Katsanos, A.J. Gorski, and 
C.J. Stephan, Phys.\ Rev.\ {\bf 182}, 1149 (1969).
\bibitem{Schiller2}A. Schiller {\it et al.}, Phys.\ Rev.\ C {\bf 68}, 054326 
(2003).
\bibitem{Tavukcu}E. Tavukcu {\it et al.}, AIP Conf.\ Proc.\ {\bf 656}, 136 
(2003).
\bibitem{Voinov}A.V. Voinov {\it et al.}, Phys.\ Rev.\ C {\bf 74}, 014314 
(2006).
\bibitem{Cwoik}S. Cwiok, J. Dudek, W. Nazarewicz, J. Skalski, and T. Werner, 
Comput.\ Phys.\ Commun.\ {\bf 46}, 379 (1987).
\bibitem{Canosa}N. Canosa, R. Rossignoli, and P. Ring, Phys. Rev.\ C {\bf 59}, 
185 (1999).
\bibitem{Liu}S. Liu and Y. Alhassid, Phys.\ Rev.\ Lett.\ {\bf 87}, 022501 
(2001).
\bibitem{Bondarenko}V.A. Bondarenko, J. Honzatko, V.A. Khitrov, Li Chol, Yu.E. 
Loginov, S.Eh.\ Malyutenkova, A.M. Sukhovoj, and I. Tomandl, in \it Proceedings
of the XII International Seminar on Interaction of Neutrons with Nuclei, Dubna,
Russia, 2004\rm, p.\ 38; nucl-ex/040630.
\bibitem{Chankova}R. Chankova {\it et al.}, Phys.\ Rev.\ C {\bf 73}, 034311 
(2006).
\bibitem{Kaneko1}K. Kaneko and M. Hasegawa, Nucl.\ Phys.\ A {\bf 740}, 95 
(2004).
\bibitem{Kaneko2}K. Kaneko and M. Hasegawa, Phys.\ Rev.\ C {\bf 72}, 024307 
(2005).
\bibitem{Schiller}A. Schiller, A. Bjerve, M. Guttormsen, M. Hjorth-Jensen, F. 
Ingebretsen, E. Melby, S. Messelt, J. Rekstad, S. Siem, and S.W. 
{\O}deg{\aa}rd, Phys.\ Rev.\ C {\bf 63}, 021306(R) (2001).
\bibitem{Melby}E. Melby, L. Bergholt, M. Guttormsen, M. Hjorth-Jensen, F. 
Ingebretsen, S. Messelt, J. Rekstad, A. Schiller, S. Siem, and S.W. 
{\O}deg{\aa}rd, Phys.\ Rev.\ Lett.\ {\bf 83}, 3150 (1999).
\bibitem{Kaneko3}K. Kaneko {\it et al.}, Phys.\ Rev.\ C {\bf 74}, 024325 
(2006).
\bibitem{Esashika}K. Esashika, H. Nakada, and K. Tanabe, Phys.\ Rev.\ C {\bf 
72}, 044303 (2005).
\end{thebibliography}
\end{document}